\documentclass[epj]{svjour}
\usepackage{graphics}
\newcommand{\be}{\begin{eqnarray}}
\newcommand{\ee}{\end{eqnarray}}
\newcommand{\bfn}{\begin{figure}[htb]\begin{center}}
\newcommand{\efn}{\end{center}\end{figure}}
\newcommand{\bfw}{\begin{figure*}[htb]\begin{center}}
\newcommand{\efw}{\end{center}\end{figure*}}
\newcommand{\btn}{\begin{table}[htb]\begin{center}}
\newcommand{\etn}{\end{center}\end{table}}
\newcommand{\btw}{\begin{table*}[htb]\begin{center}}
\newcommand{\etw}{\end{center}\end{table*}}
\newcommand{\av}[1]{\langle #1 \rangle}
\newcommand{\mapright}[1]{\smash{\mathop{\hbox to 1cm{\rightarrowfill}}\limits^{#1}}}
\newcommand{\lw}[1]{\smash{\lower 1.5ex\hbox{#1}}}
\begin{document}

\title{Analyses of multiplicity distributions with $\eta_c$ and Bose-Einstein correlations at LHC by means of generalized Glauber-Lachs formula}
\author{Takuya Mizoguchi\inst{1}\thanks{mizoguti@toba-cmt.ac.jp} \and Minoru Biyajima\inst{2}\thanks{biyajima@azusa.shinshu-u.ac.jp}
}                     
\offprints{}          
\institute{Toba National College of Maritime Technology, Toba 517-8501, Japan \and Department of Physics and College of General Education, Shinshu University, Matsumoto 390-8621, Japan}
\date{Received: date / Revised version: date}
%
\abstract{
Using the negative binomial distribution (NBD) and the generalized Glauber-Lachs (GGL) formula, we analyze the data on charged multiplicity distributions with pseudo-rapidity cutoffs $\eta_c$ at 0.9, 2.36, and 7 TeV by ALICE Collaboration and at 0.2, 0.54, and 0.9 TeV by UA5 Collaboration. We confirm that the KNO scaling holds among the multiplicity distributions with $\eta_c =$ 0.5 at $\sqrt{s} =$ 0.2$\sim$2.36 TeV and estimate the energy dependence of a parameter $1/k$ in NBD and parameters $1/k$ and $\gamma$ (the ratio of the average value of the coherent hadrons to that of the chaotic hadrons) in the GGL formula. Using empirical formulae for the parameters $1/k$ and $\gamma$ in the GGL formula, we predict the multiplicity distributions with $\eta_c =$ 0.5 at 7 and 14 TeV. Data on the 2nd order Bose-Einstein correlations (BEC) at 0.9 TeV by ALICE Collaboration and 0.9 and 2.36 TeV by CMS Collaboration are also analyzed based on the GGL formula. Prediction for the 3rd order BEC at 0.9 and 2.36 TeV are presented. Moreover, the information entropy is discussed.
} 
\maketitle

\section{Introduction}
\label{se_01}
Very recently ALICE Collaboration has investigated the multiplicity distributions with pseudo-rapidity cutoffs ($\eta_c =$ 0.5, 1.0, and 1.3) at $\sqrt{s} =$ 0.9 and 2.36 TeV \cite{Aamodt:2010ft}. Therein, it has compared its data with the data at 0.2, 0.54, and 0.9 TeV by UA5 Collaboration \cite{Alner:1985zc,Ansorge:1988kn}, and concluded that the combined data with $\eta_c =$ 0.5 at 0.2, 0.9, and 2.35 TeV are fairly well described by the single NBD (negative binomial distribution)~\cite{GrosseOetringhaus:2009kz,Fuglesang:1990aa}. Moreover, ALICE Collaboration has reported that the KNO scaling~\cite{Koba:1972ng} holds among the combined data with $\eta_c =$ 0.5 at 0.2, 0.9, and 2.35 TeV. The first aim of this paper is to confirm the statement above mentioned in~\cite{Aamodt:2010ft} and to analyze the same data by the GGL(generalized Glauber-Lachs) formula~\cite{Biyajima:1982un,Biyajima:1984aq}. To investigate possibility of predictions on multiplicity distributions with $\eta_c =$ 0.5 at 7 and 14 TeV~\cite{Aamodt:2010pp}, empirical formulae (for $1/k$ and $\gamma$) are adopted.

Moreover, ALICE and CMS Collaborations have reported the data on Bose-Einstein correlations (BEC) at 0.9 and 2.36 TeV~\cite{Aamodt:2010jj,Khachatryan:2010un}. Thus we are going to investigated them based on a conventional formula with the degree of coherence and the GGL formula.

First of all, the NBD is introduced in the following:
\be
P_k(n) = \frac{\Gamma(n+k)}{\Gamma(n+1)\Gamma(k)} 
     \frac{(\av{n}/k)^n}{(1+\av{n}/k)^{n+k}},
\label{eq_01}
\ee
where $\av{n}$ and $k$ are the average multiplicity and the intrinsic parameter, respectively. In the KNO scaling limit ($n$ and $\av{n}$ are large, but the ratio $z = n/\av{n}$ is finite), for the quantity $\av{n}P(n,\:\av{n})$ the following gamma distribution is derived from Eq.~(\ref{eq_01}) as
\be 
\psi_k(z) = \frac{k^k}{\Gamma(k)}z^{k-1}e^{-kz}
\label{eq_02}
\ee
Second we turn to the GGL formula which is expressed as follows:
\be
P_k(n) &=& \frac{(p\av{n}/k)^n}{(1+p\av{n}/k)^{n+k}}
     \exp\left[-\frac{\gamma p\av{n}}{1+p\av{n}/k}\right]\nonumber\\
     &&\cdot L_n^{(k-1)}\left(-\frac{\gamma k}{1+p\av{n}/k}\right),
\label{eq_03}
\ee
where $\gamma = |\zeta|^2/A$ (the ratio of the ratio of the average value of the coherent hadrons to that of the chaotic hadrons), $p = 1/(1+\gamma )$, and $L_n^{(k-1)}$ stands for the Laguerre polynomials, respectively. Here it should be stressed that the GGL formula has three limits, the original GL formula ($k=1$)~\cite{Glauber:1965aa}, the NBD ($\gamma = 0$), and the Poisson distribution ($k =$ 1 and $\gamma = \infty$, or $\gamma =$ 0 and $k = \infty$). 

\setlength{\unitlength}{0.7mm}
\begin{picture}(110,45)(5,-2)
\put(0,20){\makebox(20,10)[r]{Eq.~(\ref{eq_03})}}
\put(44,30){\makebox(60,10)[l]{($k=$ 1) original GL}}
\put(44,20){\makebox(60,10)[l]{($\gamma = 0$) NBD}}
\put(44,0){\makebox(60,10)[l]{($k=$ 1, $\gamma = \infty$) Poisson distribution}}
\put(64,10){\makebox(60,10)[l]{($k= \infty$)}}
\put(23,28){\vector(3,1){18}}
\put(23,25){\vector(1,0){18}}
\put(23,22){\vector(4,-3){18}}
\put(60,20){\vector(0,-1){10}}
\end{picture}

The KNO scaling function of Eq.~(\ref{eq_03}) is given in the following
\be 
&&\psi_k(z,\:p) = \left(\frac{k}{p}\right)^k 
        \left[\frac{z}{\sqrt{z(k/p)^2(1-p)}}\right]^{k-1} \nonumber\\
        &&\quad \cdot\exp\left[-\frac{k}{p}(1-p+z)\right] 
        I_{k-1}\left(2\sqrt{z(k/p)^2(1-p)}\right)
\label{eq_04}
\ee
where $I_{k-1}$ is the modified Bessel function. Eq.~(\ref{eq_04}) becomes the gamma distribution, as $\gamma =$ 0.

In order to analyze of Bose-Einstein correlations (GGLP effect~\cite{Goldhaber:1960sf}, or hadronic HBT effect~\cite{Biyajima:1990ku,Weiner:1997kg}) at LHC, we are going to use the following formulae: The first one is well known as the conventional formula, 
\be
N^{(--)}/N^{BG} ({\rm conventional\ formula})
= 1 + \lambda\: E_{2B}^2,
\label{eq_05}
\ee
where the parameter $\lambda$ is named the degree of coherence, and $E_{2B}$ is function of momentum transfer ($Q^2 = -(p_1-p_2)^2$) and the range of interaction $R$. $E_{2B} = \exp(-R^2Q^2)$ (Gaussian formula) and/or $E_{2B} = \exp(-R\sqrt{Q^2})$ (exponential formula) are used. The second one proposed in Ref.~\cite{Biyajima:1990ku} is relating to the GGL formula as follows
\be
&& N^{(--)}/N^{BG} ({\rm (GGL)})
=  1 + [2p(1-p)E_{2B} + p^2E_{2B}^2]/k \nonumber\\
&& \hspace{22mm}\mapright{k\to 1}\ 1 + [2p(1-p)E_{2B} + p^2E_{2B}^2],
\label{eq_06}
\ee
where $p = 1/(1 + \gamma)$, and $(k\to 1)$ means the identical charged ensemble.

The present paper is organized in the following: In the second paragraph, we analyze the data with pseudo-rapidity cutoffs by means of the NBD and the GGL formula. In the third paragraph, the distributions of the KNO scaling are analyzed by Eqs.~(\ref{eq_02}) and (\ref{eq_04}). In the fourth paragraph, we consider the physical meaning of the parameter $\gamma$ in the GGL formula: Data on the 2nd order BEC at 0.9 and 2.36 TeV~\cite{Aamodt:2010jj,Khachatryan:2010un} are analyzed by Eqs.~(\ref{eq_05}) and (\ref{eq_06})~\cite{Biyajima:1990ku,Weiner:1997kg}. In last paragraph, concluding remarks and discussion for the information entropy are given.

\section{Analyses of data on multiplicity distributions by the NBD and the GGL formula}
\label{se_02}
Utilizing Eqs.~(\ref{eq_01}) and (\ref{eq_03}), we analyze the data with pseudo-rapidity cutoffs ($\eta_c =$ 0.5, 1.0, and 1.3) at 0.2, 0.54, 0.9 and 2.36 TeV. Hereafter we use the CERN-MINUIT program. Results are shown in Fig.~\ref{fi_01} and Table~\ref{ta_01}.
\bfw
\resizebox{0.45\textwidth}{!}{\includegraphics{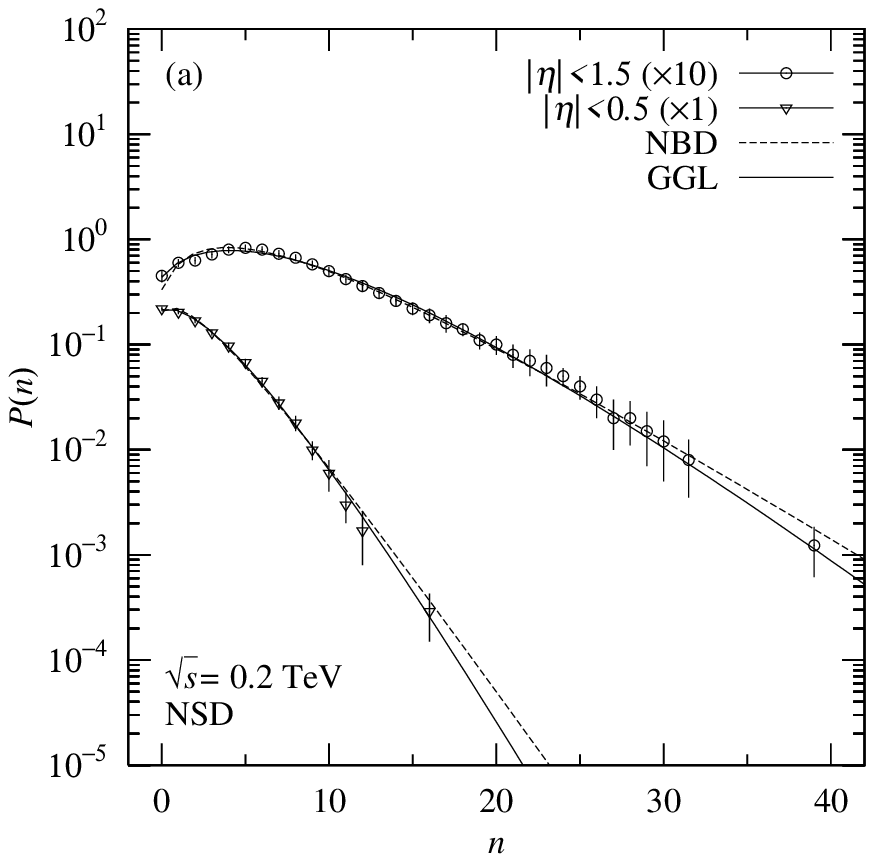}}
\resizebox{0.45\textwidth}{!}{\includegraphics{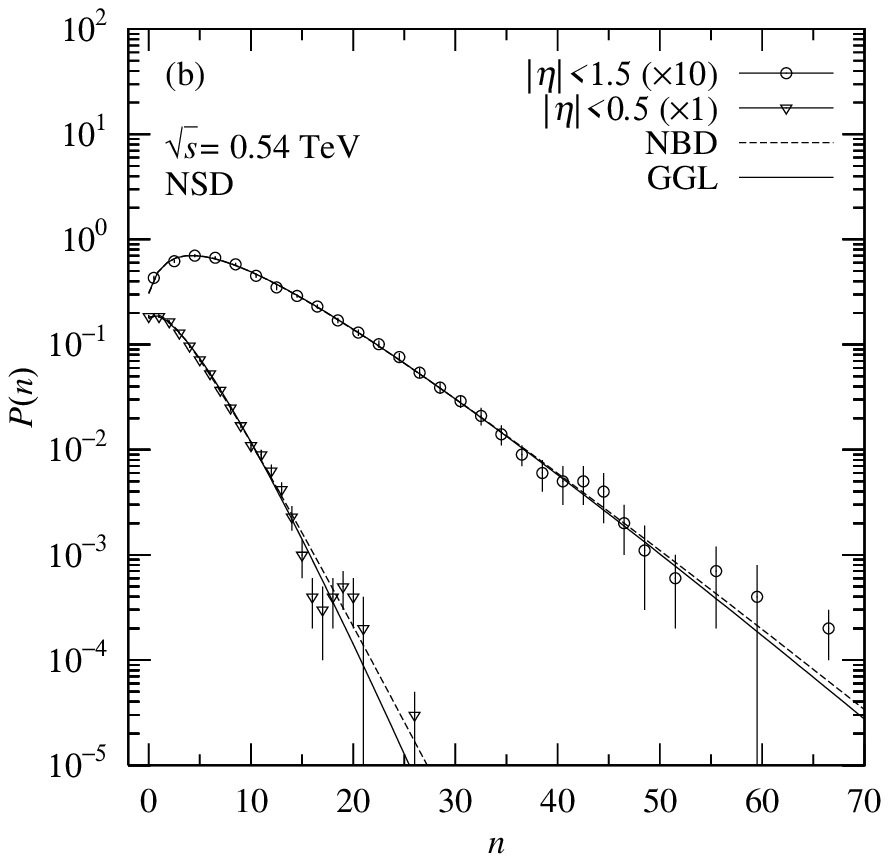}}\\
\resizebox{0.45\textwidth}{!}{\includegraphics{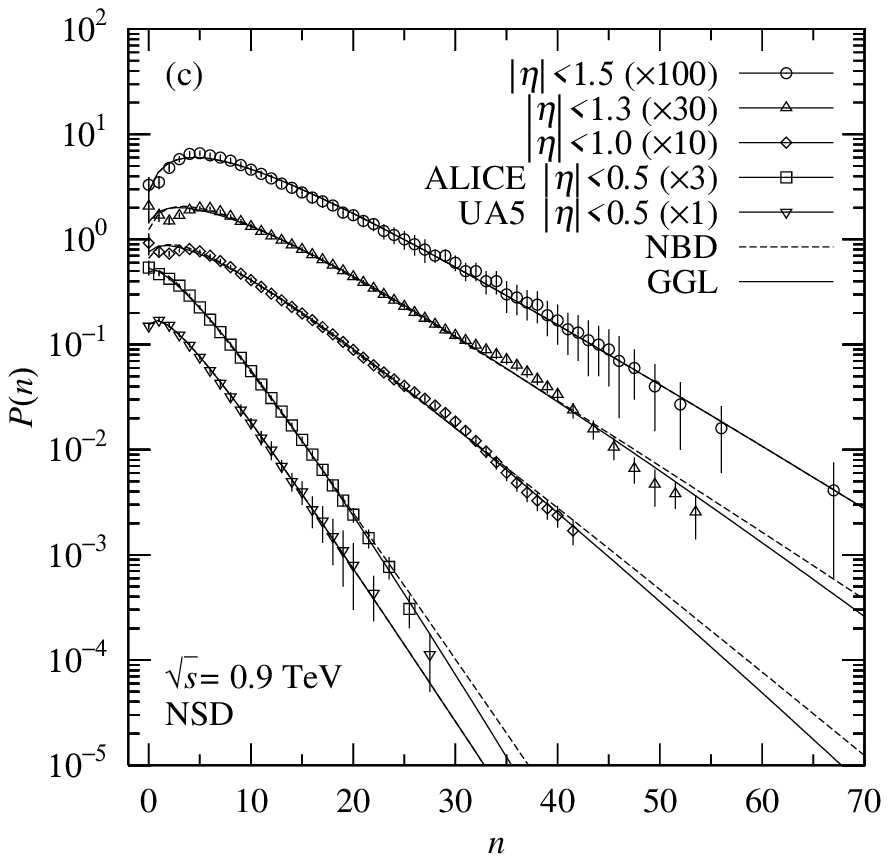}}
\resizebox{0.45\textwidth}{!}{\includegraphics{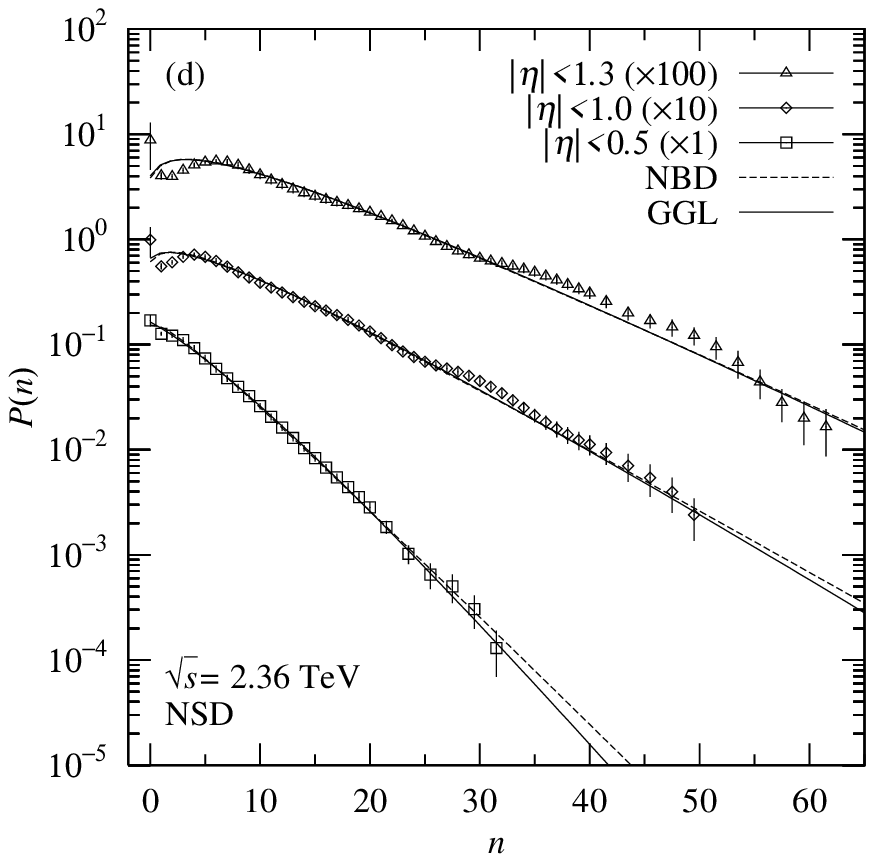}}
\resizebox{0.45\textwidth}{!}{\includegraphics{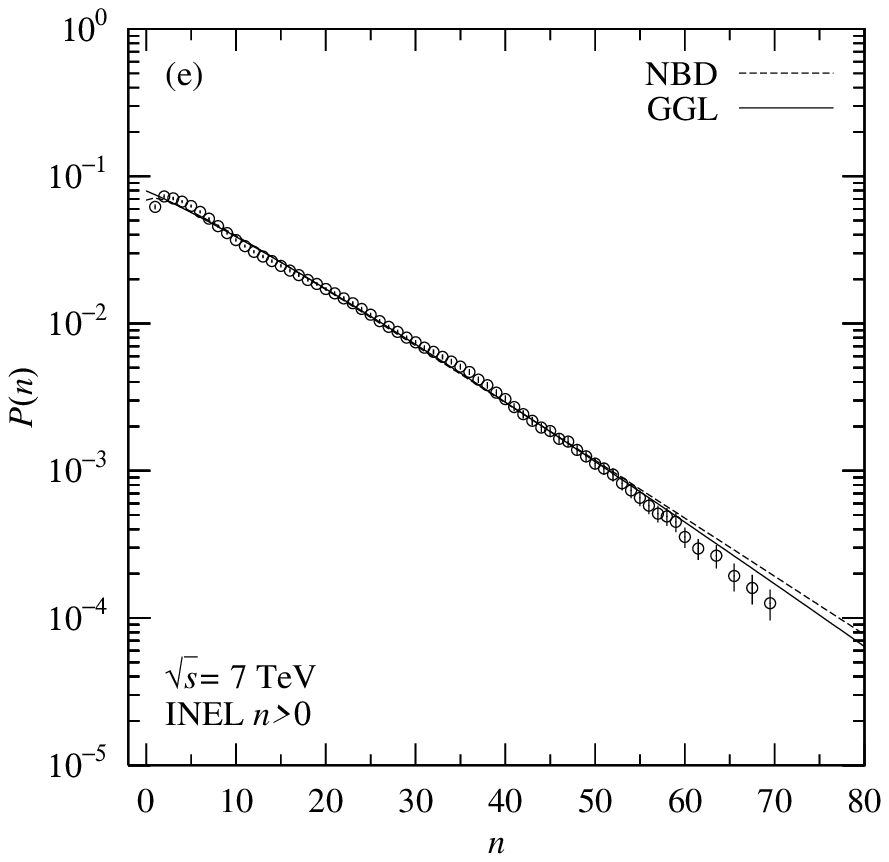}}
\caption{Analyses of data with $|\eta|<\eta_c$ by means of Eqs.~(\ref{eq_01}) and (\ref{eq_03}).}
\label{fi_01}
\efw
\btw
\caption{Results in analyses of multiplicity distributions (NSD except for 7 TeV).}
\label{ta_01}    
\vspace{2mm}
\begin{tabular}{c|cc|ccc}
\hline
& \multicolumn{2}{c|}{NBD} & \multicolumn{3}{c}{GGL}\\
$\eta_c$ & $1/k$ & $\chi^2/$NDF & $1/k$ & $\gamma$ & $\chi^2/$NDF\\
\hline
\multicolumn{6}{c}{UA5, $\sqrt{s} =$ 0.2 TeV}\\
\hline
0.5 & 0.56$\pm$0.03 & 11.5/13 & 1.00$\pm$0.22 &  2.2$\pm$0.2  & 3.8/12\\
1.5 & 0.45$\pm$0.02 & 19/32   & 0.89$\pm$0.46 &  2.5$\pm$2.2  & 10.7/31\\
\hline
\multicolumn{6}{c}{UA5, $\sqrt{s} =$ 0.54 TeV}\\
\hline
0.5 & 0.59$\pm$0.02 & 30/22   & 0.86$\pm$0.19 &  1.4$\pm$0.7  & 21/21\\
1.5 & 0.50$\pm$0.01 & 13.8/28 & 0.52$\pm$0.03 & 0.22$\pm$0.26 & 13.5/27\\
\hline
\multicolumn{6}{c}{UA5, $\sqrt{s} =$ 0.9 TeV}\\
\hline
0.5 & 0.65$\pm$0.03 & 2.1/22  & 0.65$\pm$0.03 &  0.0$\pm$11.3 & 2.1/21\\
1.5 & 0.56$\pm$0.02 & 18.7/51 & 0.56$\pm$0.02 &  0.0$\pm$63.0 & 18.7/50\\
\hline
\multicolumn{6}{c}{ALICE, $\sqrt{s} =$ 0.9 TeV}\\
\hline
0.5 & 0.69$\pm$0.01 & 10.4/23 & 0.89$\pm$0.13 & 0.88$\pm$0.42 & 3.7/22\\
1.0 & 0.65$\pm$0.01 & 44/41   & 0.79$\pm$0.08 & 0.74$\pm$0.28 & 35/40\\
1.3 & 0.61$\pm$0.01 & 70/47   & 0.74$\pm$0.08 & 0.70$\pm$0.29 & 62/46\\
\hline
\multicolumn{6}{c}{ALICE, $\sqrt{s} =$ 2.36 TeV}\\
\hline
0.5 & 0.82$\pm$0.02 & 19/26   & 1.00$\pm$0.11 & 0.75$\pm$0.08 & 14/25\\
1.0 & 0.73$\pm$0.02 & 82/45   & 0.79$\pm$0.09 & 0.41$\pm$0.33 & 80/44\\
1.3 & 0.67$\pm$0.02 & 145/51  & 0.70$\pm$0.04 & 0.25$\pm$0.20 & 144/50\\
\hline
\multicolumn{6}{c}{ALICE, $\sqrt{s} =$ 7 TeV (inelastic, $n>0$)}\\
\hline
1.0 & 0.88$\pm$0.01 & 217/62   & 0.99$\pm$0.03 & 0.45$\pm$0.08 & 191/61\\
& \multicolumn{2}{c|}{$\av{n} =$ 11.6$\pm$0.1, $c =$ 1.07$\pm$0.01} 
& \multicolumn{3}{c}{$\av{n} =$ 11.5$\pm$0.1, $c =$ 1.08$\pm$0.01}\\
\hline
\end{tabular}
\etw
From results based on analyses of the NBD, we observe that $1/k$ increases gradually as $\sqrt{s}$ increases. On the other hand, the estimated sets of ($1/k$ and $\gamma$ ) in the GGL formula show different behavior. Using the results of parameters for data with $\eta_c =$ 0.5, in particular, we estimate the energy dependence of $1/k$ in the NBD, and those of $1/k$ and $\gamma$ in the GGL formula as follows:
\be
1/k^{\rm (NBD)} &=& 0.49 + 0.12 \ln (\sqrt{s}/0.2),
\label{eq_07}\\
1/k^{\rm (GGL)} &=& 0.89 + 0.031 \ln (\sqrt{s}/0.2),
\label{eq_08}\\
\gamma^{\rm (MD)} &=& 2.10 - 0.55 \ln (\sqrt{s}/0.2),
\label{eq_09}
\ee
where MD stands for the multiplicity distribution. Energy dependences of parameters $1/k^{\rm (NBD)}$, $1/k^{\rm (GGL)}$ and $\gamma^{\rm (MD)}$ are shown in Fig.~\ref{fi_02}.

Moreover, the data with $\eta_c =$ 1.0 at 7 TeV have been reported by ALICE Collaboration~\cite{Aamodt:2010pp}. The data are relating to inelastic events with a positive condition INEL$>0$ and no value on $\av{n}$, nevertheless. Then in our analyses, the following modifications with a normalization factor $c$ is adopted:
\be
{\rm Eq.}~(\ref{eq_01})\to c\times {\rm Eq.}~(\ref{eq_01})\ ({\rm with~free}\ \av{n}),
\label{eq_10}\\
{\rm Eq.}~(\ref{eq_02})\to c\times {\rm Eq.}~(\ref{eq_02})\ ({\rm with~free}\ \av{n}),
\label{eq_11}
\ee
The estimated values are added in Table~\ref{ta_01}. It can be said that multiplicity distribution at 7 TeV is fairly well explained by the GGL formula as well as NBD. (See Fig.~\ref{fi_01}.)
\bfn
\resizebox{0.48\textwidth}{!}{\includegraphics{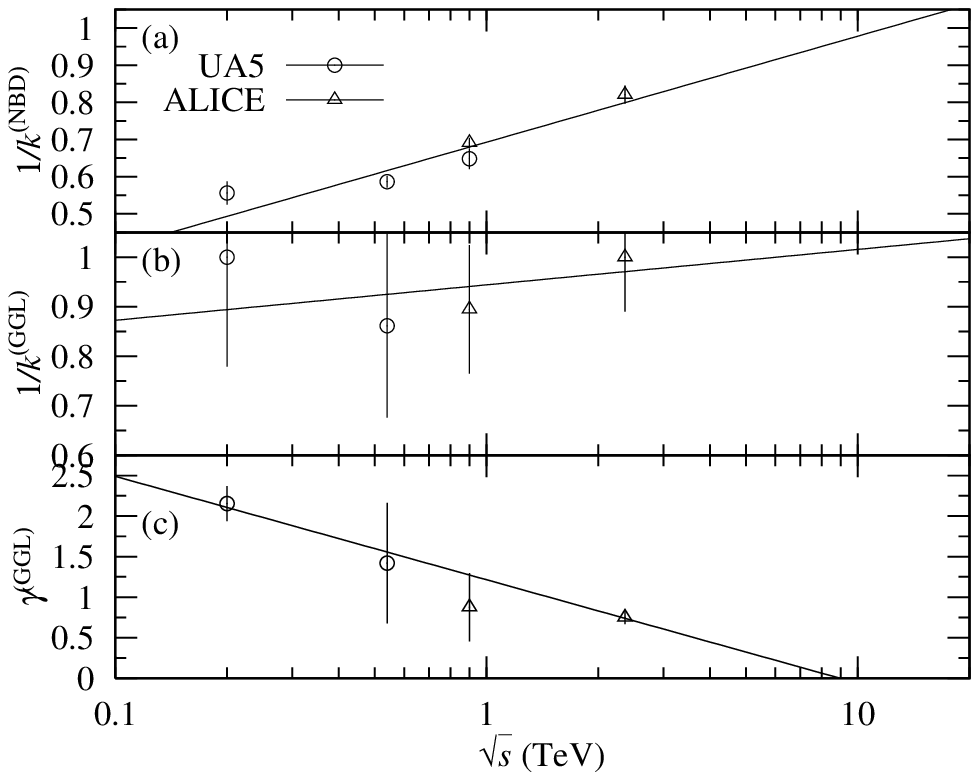}}
\label{fi_02}
\caption{Energy dependence of parameters $1/k^{\rm (NBD)}$, $1/k^{\rm (GGL)}$, and $\gamma^{(MD)}$ for data with $\eta_c =$ 0.5. Values at 0.9 TeV by UA5 Collaboration are omitted in GGL formula, because of extreme error bars.}
\efn

To get more useful physical information on multiplicity distributions, the KNO scaling distributions are investigated in the next paragraph.

\btn 
\caption{Energy dependence of parameters ($1/k^{\rm (NBD)}$, $1/k^{\rm (GGL)}$, and $\gamma^{\rm (MD)}$) for multiplicity distributions with $\eta_c =$ 0.5. Eqs.~(\ref{eq_07}), (\ref{eq_08}), and (\ref{eq_09}) are used. See Fig.~\ref{fi_02}.}
\label{ta_02}
\vspace{2mm}
\begin{tabular}{cccc}
\hline
$\sqrt{s}$ & $1/k^{\rm (NBD)}$ & $1/k^{\rm (GGL)}$ & $\gamma^{\rm (MD)}$\\
\hline
0.9  & 0.69 &  0.89 & 0.88\\
2.36 & 0.82 &  1.00 & 0.75\\
\hline
\multicolumn{4}{c}{calculated values}\\
7    & 0.93 &  1.0 & 0.14\\
14   & 1.00 &  1.0 & 0.0\\
\hline
\end{tabular}
\etn

\section{Analyses of data on KNO scaling distributions by Eqs.~(\ref{eq_02}) and (\ref{eq_04})}
\label{se_03}
Utilizing the KNO scaling variable $z (= n/\av{n})$, data on the KNO scaling distributions $\av{n} P(n,\: \av{n})$ are shown in Fig.~\ref{fi_03}. Our results analyzed by Eqs.~(\ref{eq_02}) and (\ref{eq_04}) are given in Table~\ref{ta_03}.
\bfw
\resizebox{0.45\textwidth}{!}{\includegraphics{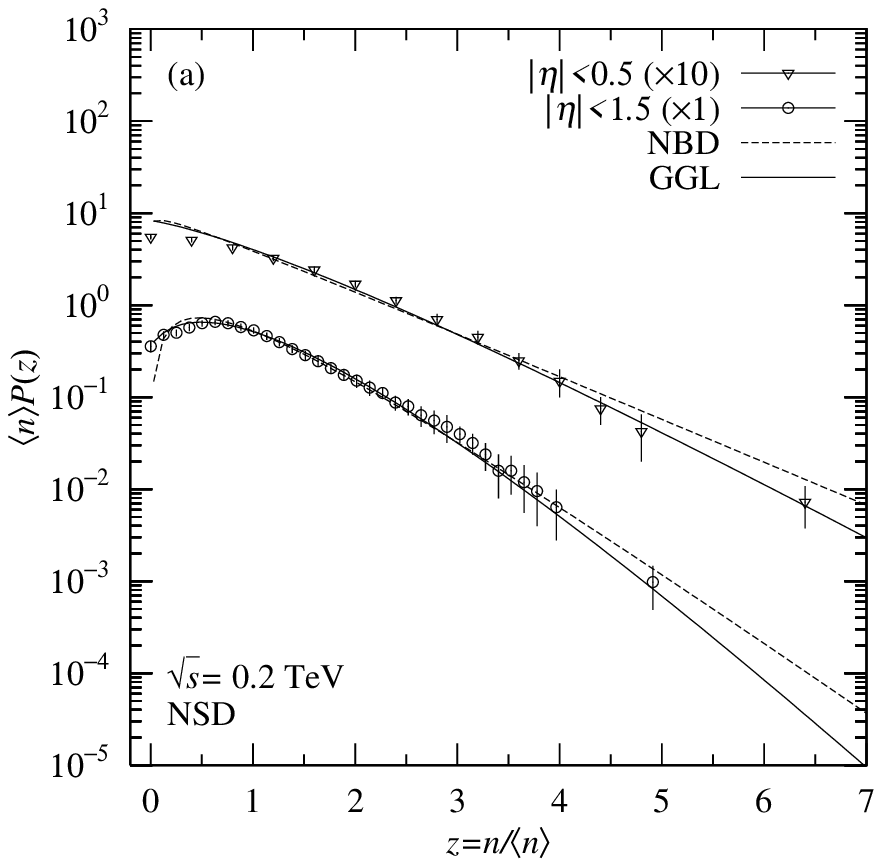}}
\resizebox{0.45\textwidth}{!}{\includegraphics{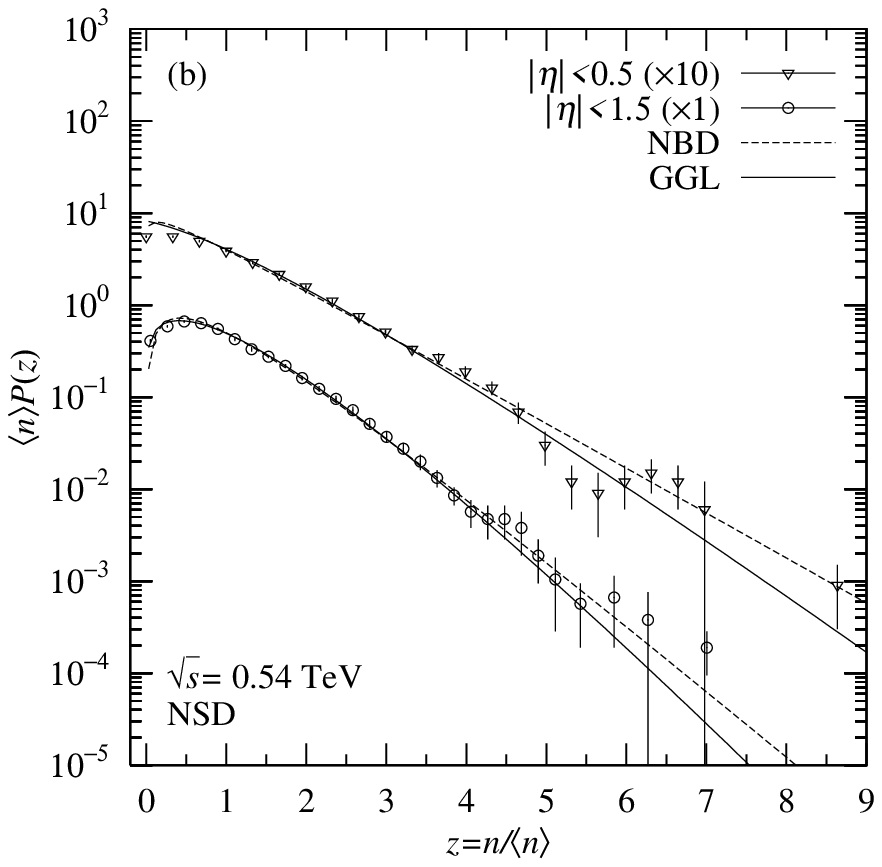}}\\
\resizebox{0.45\textwidth}{!}{\includegraphics{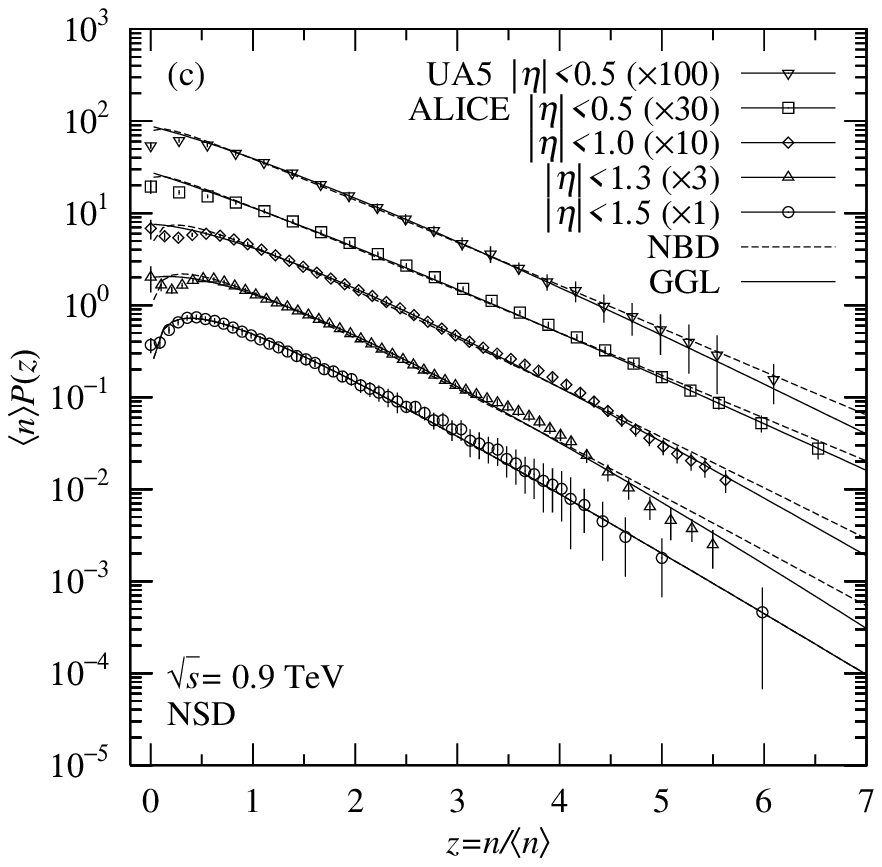}}
\resizebox{0.45\textwidth}{!}{\includegraphics{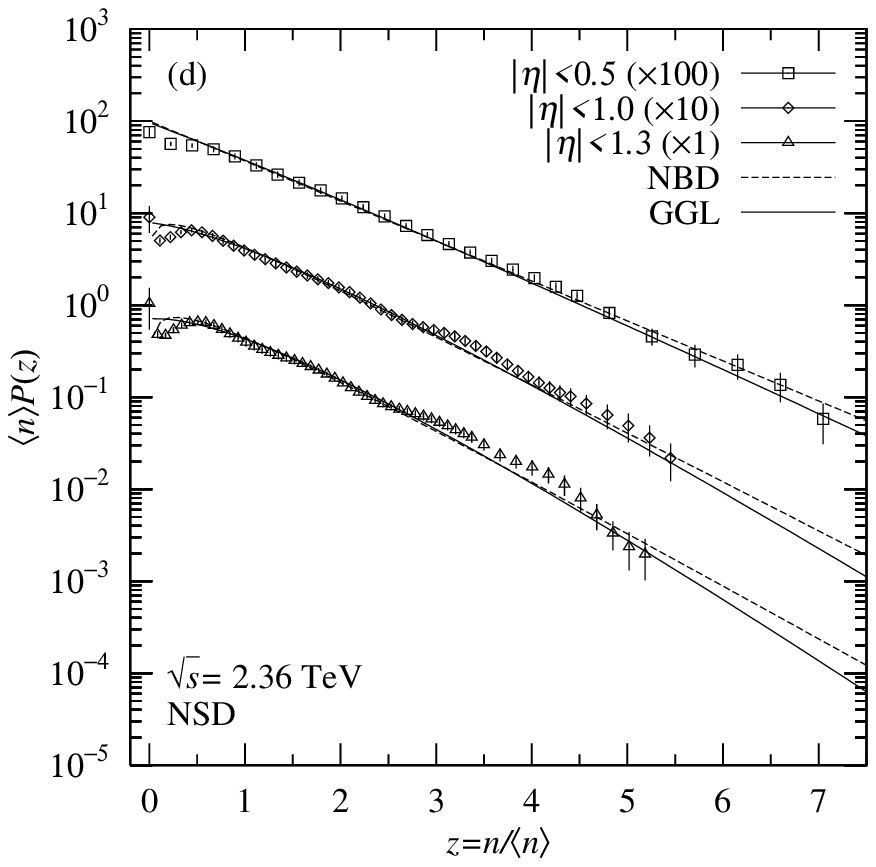}}\\
\resizebox{0.45\textwidth}{!}{\includegraphics{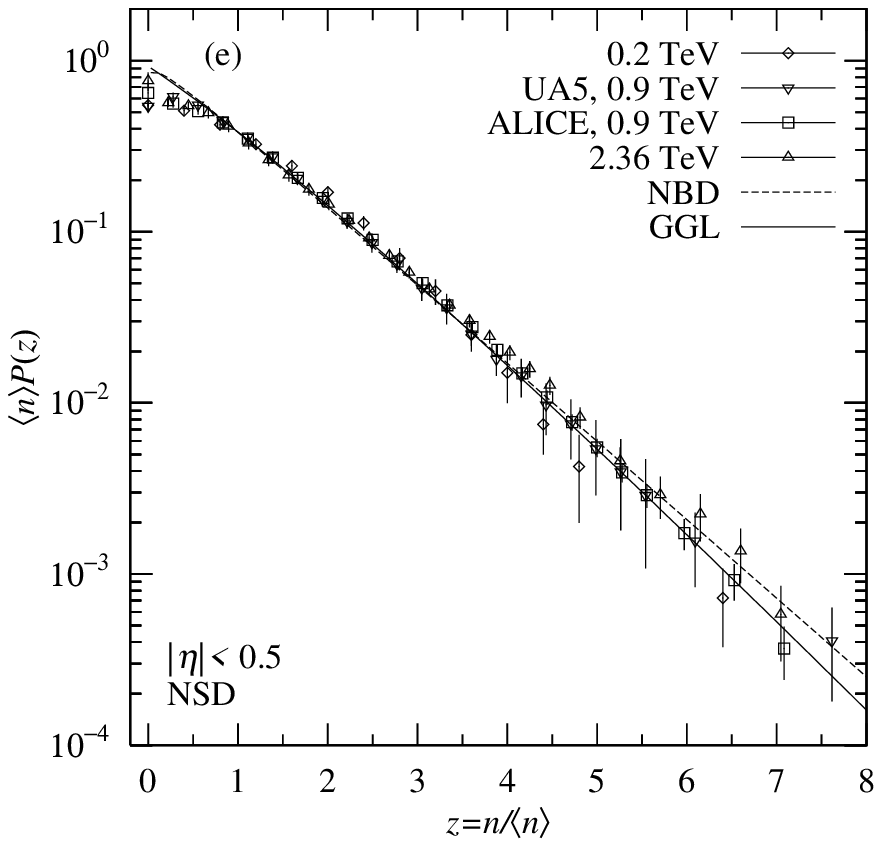}}
\caption{Analyses of KNO scaling distributions $\av{n}P(n)$'s. The same data of Fig.~\ref{fi_01} are described by KNO scaling variable $z=n/\av{n}$. Eqs.~(\ref{eq_02}) and (\ref{eq_04}) are used.}
\label{fi_03}
\efw
\btw
\caption{Results in analyses of KNO scaling distributions $\av{n}P(n)$ described by KNO variable $z = n/\av{n}$.}
\label{ta_03}    
\vspace{2mm}
\begin{tabular}{c|cc|ccc}
\hline
& \multicolumn{2}{c|}{NBD} & \multicolumn{3}{c}{GGL}\\
$\eta_c$ & $1/k$ & $\chi^2/$NDF & $1/k$ & $\gamma$ & $\chi^2/$NDF\\
\hline
\multicolumn{6}{c}{UA5, $\sqrt{s} =$ 0.2 TeV}\\
\hline
0.5 & 0.92$\pm$0.04 & 59/12  & 1.00$\pm$0.01 & 0.72$\pm$0.12 & 38/11\\
1.5 & 0.53$\pm$0.02 & 28/31  & 1.00$\pm$0.13 &  2.2$\pm$0.1  & 14/30\\
\hline
\multicolumn{6}{c}{UA5, $\sqrt{s} =$ 0.54 TeV}\\
\hline
0.5 & 0.87$\pm$0.02 & 172/21 & 1.00$\pm$0.00 & 0.77$\pm$0.06 & 100/20\\
1.5 & 0.58$\pm$0.01 & 41/28  & 0.75$\pm$0.04 & 0.93$\pm$0.18 & 24/27\\
\hline
\multicolumn{6}{c}{UA5, $\sqrt{s} =$ 0.9 TeV}\\
\hline
0.5 & 0.91$\pm$0.04 & 22/21  & 1.00$\pm$0.02 & 0.59$\pm$0.11 & 13/20\\
1.5 & 0.62$\pm$0.02 &15.3/50 & 0.62$\pm$0.02 &  0.0$\pm$17.0 & 15.3/49\\
\hline
\multicolumn{6}{c}{ALICE, $\sqrt{s} =$ 0.9 TeV}\\
\hline
0.5 & 0.92$\pm$0.01 & 80/22  & 1.00$\pm$0.01 & 0.44$\pm$0.04 & 54/21\\
1.0 & 0.76$\pm$0.01 & 99/40  & 1.00$\pm$0.01 & 0.93$\pm$0.04 & 56/39\\
1.3 & 0.69$\pm$0.01 & 110/46 & 1.00$\pm$0.03 & 1.20$\pm$0.04 & 73/45\\
\hline
\multicolumn{6}{c}{ALICE, $\sqrt{s} =$ 2.36 TeV}\\
\hline
0.5 & 1.00$\pm$0.05 & 94/25  & 1.00$\pm$0.01 & 0.31$\pm$0.08 & 87/24\\
1.0 & 0.79$\pm$0.02 & 128/44 & 1.00$\pm$0.03 & 0.85$\pm$0.06 & 118/43\\
1.3 & 0.73$\pm$0.02 & 159/50 & 1.00$\pm$0.09 & 1.07$\pm$0.06 & 156/49\\
\hline\hline
\multicolumn{6}{c}{combined UA5, 0.2, 0.9 TeV, ALICE, 0.9, 2.36 TeV}\\
\hline
0.5 & 0.94$\pm$0.01 & 264/83 & 1.00$\pm$0.00 & 0.44$\pm$0.03 & 203/82\\
\hline
\end{tabular}
\etw

Second, we combine the data with $\eta_c =$ 0.5 at 0.2, 0.9 and 2.36 TeV and analyze them by Eq.~(\ref{eq_02}) (the gamma distribution) and Eq.~(\ref{eq_04}) (the modified Bessel function). Since data with $\eta_c =$ 0.5 at 0.54 TeV are not well explained by two formulae, three data are regarded as exceptional among them. In the final paragraph, we consider this fact. Results on data with $\eta_c =$ 1.3 at 2.36 TeV show large $\chi^2$ values; Large values of $\chi^2$'s ($\chi^2 >>$ NDF) denote that the single NBD or the single GGL formula should be improved for explanations of data. See, for example, Ref.~\cite{Giovannini:1998zb}.

Moreover, the results on $1/k$ in Table~\ref{ta_03} by Eq.~(\ref{eq_02}) (gamma distribution) show similar behavior to those in Table~\ref{ta_01}. On the other hand, the sets of ($1/k$ and $\gamma$) estimated by Eq.~(\ref{eq_04}) are almost approximately ($1/k \cong$ 1.0) at 0.2$\sim$2.36 TeV. The parameter $\gamma$ depends on colliding energies, which is roughly expressed as
\be
\gamma^{\rm (KNO)} &=& 0.96 - 0.29 \ln (\sqrt{s}/0.2)
\label{eq_12}
\ee
Comparing Eq.~(\ref{eq_12}) with (\ref{eq_09}), we observe that $\gamma^{\rm (KNO)} \approx \gamma^{\rm (MD)}/3$. From the bottom panel of Fig.~\ref{fi_03}, we know that coincidence among combined data with $\eta_c =$ 0.5 (the number of data points is 84) is fairly well explained  by the GGL formula with smaller value of $\chi^2$ than that of the NBD.

\section{Analyses of data on the 2nd order BEC by means of Eqs.~(\ref{eq_05}) and (\ref{eq_06})}
\label{se_04}
We analyze the data on BEC at LHC by the use of Eqs.~(\ref{eq_05}) and (\ref{eq_06}) with $E_{2B} = \exp(-R^2Q^2)$ and/or $E_{2B} = \exp(-R\sqrt{Q^2})$. The following modification, i.e., introducing the normalization factor, is used. $c$ is reflecting the long range correlation $(1+\delta Q)^{-1}$ in many data.
\be
{\rm Eq.\:}(\ref{eq_05}) &\rightarrow & c\times {\rm Eq.\:}(\ref{eq_05}) \mapright{Q\to 0} c(1 + \lambda),
\label{eq_13}\\
{\rm Eq.\:}(\ref{eq_06}) &\rightarrow & c\times {\rm Eq.\:}(\ref{eq_06}) \mapright{Q\to 0} c\left[1 + \frac{(1 + 2\gamma^{(\rm BEC)}) }{( 1 + \gamma^{(\rm BEC)})^2 }\right].
\label{eq_14}
\ee
Results are depicted in Table~\ref{ta_04} and Fig.~\ref{fi_04}. In Eq.~(\ref{eq_14}), the effective degree of coherence ``$\lambda$'' is ``$(1 + 2\gamma)/( 1 + \gamma)^2$''. In our concrete analyses, we obtained that $1/k^{\rm (BEC)} =$ 1.    Because estimated value of this seems to be reasonable for the identical charged ensemble, it is omitted. $\gamma^{\rm (BEC)}$ is similar to the value at 0.9 TeV by ALICE Collaboration in Table~\ref{ta_01}.

Furthermore, by the use of Eqs.~(\ref{eq_05}) and (\ref{eq_06}) with $c$, we have analyzed the data on BEC at 0.9 and 2.36 TeV by CMS Collaboration~\cite{Khachatryan:2010un}. Results are shown in Fig.~\ref{fi_04} and Table~\ref{ta_04}. It is emphasized that the ratio $\gamma^{\rm (BEC)}$ decreases, as the colliding energy increases. In other words, the effective degree of coherence ``$\lambda$'' and the range of interaction $R$ increases from 0.9 to 2.36 TeV. To draw more significant meaning about the parameter $\gamma$, we need the BEC measured in data with $\eta_c =$ 0.5.

It is worthwhile to predict the 3rd BEC at 0.9 TeV using the same condition with $M\le$ 6, 0.1 $\le k_T \le$ 0.55 GeV/$c$. Using estimated values of $\gamma^{\rm (BEC)}$ and $R$ in the 2nd BEC by ALICE Collaboration, we can predict the 3rd order BEC; The following formula~\cite{Biyajima:1990ku} is used,
\be
N^{(3-)}/N^{BG}
&=& 1 + 6p(1-p)\exp\left(-\frac{1}{3}R\sqrt{Q_3^2}\right)\nonumber\\ 
&&+ 3p^2(3-2p)\exp\left(-\frac{2}{3}R\sqrt{Q_3^2}\right)\nonumber\\
&&+ 2p^3\exp\left(-R\sqrt{Q_3^2}\right),
\label{eq_15}
\ee
where $p=1/(1+\gamma)$ and $Q_3^2 = Q_{12}^2 + Q_{23}^2 + Q_{31}^2$. Our prediction is depicted in Fig.~\ref{fi_04}(d).

Furthermore, predictions on the 3rd order BEC at 0.9 and 2.36 TeV for CMS Collaboration are also displayed in Fig.~\ref{fi_04}(d) and (e).

\bfw
\resizebox{0.40\textwidth}{!}{\includegraphics{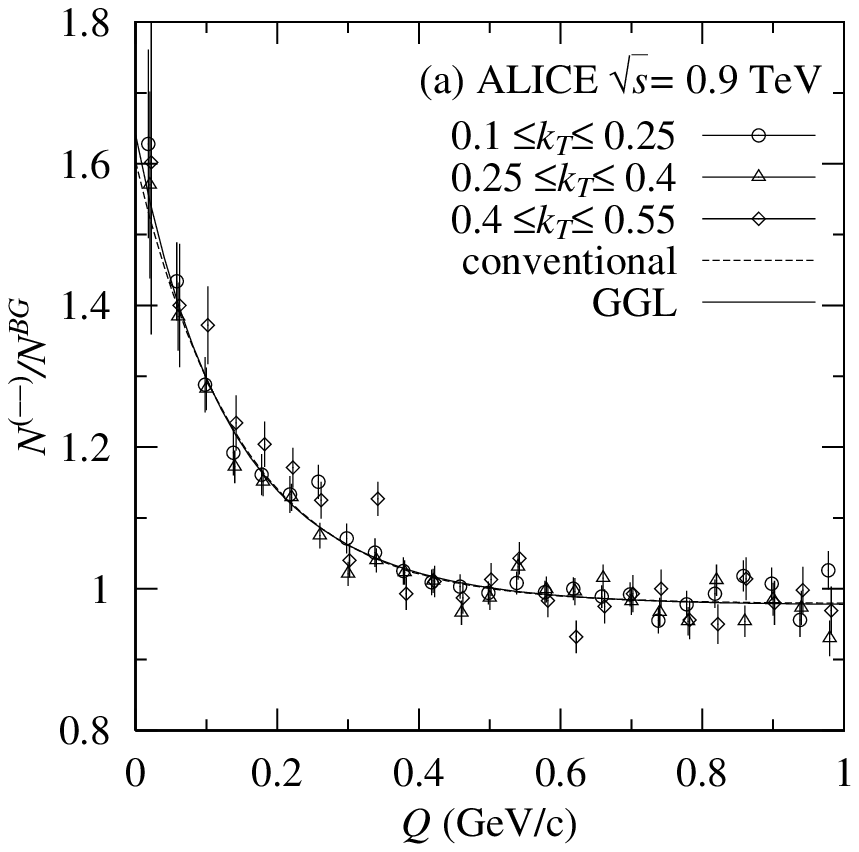}}
\resizebox{0.56\textwidth}{!}{\includegraphics{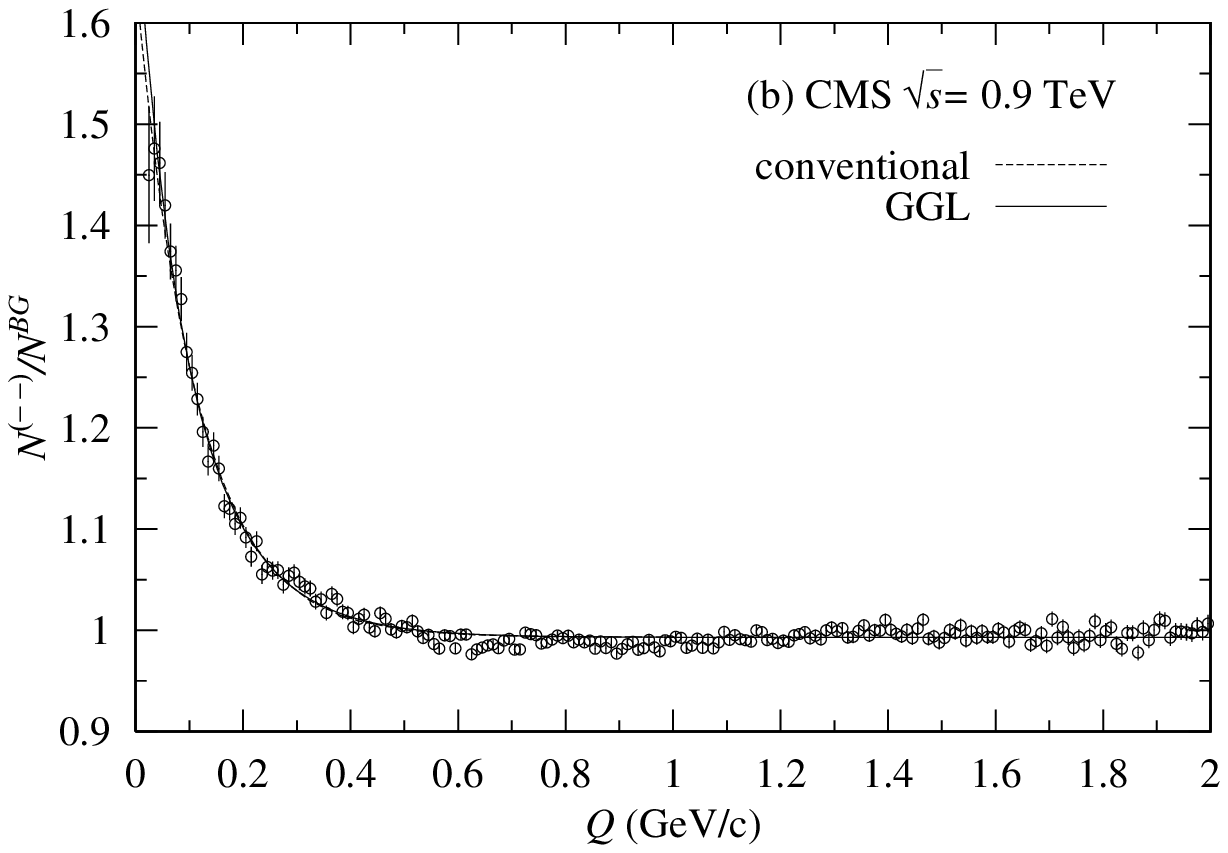}}\\
\resizebox{0.56\textwidth}{!}{\includegraphics{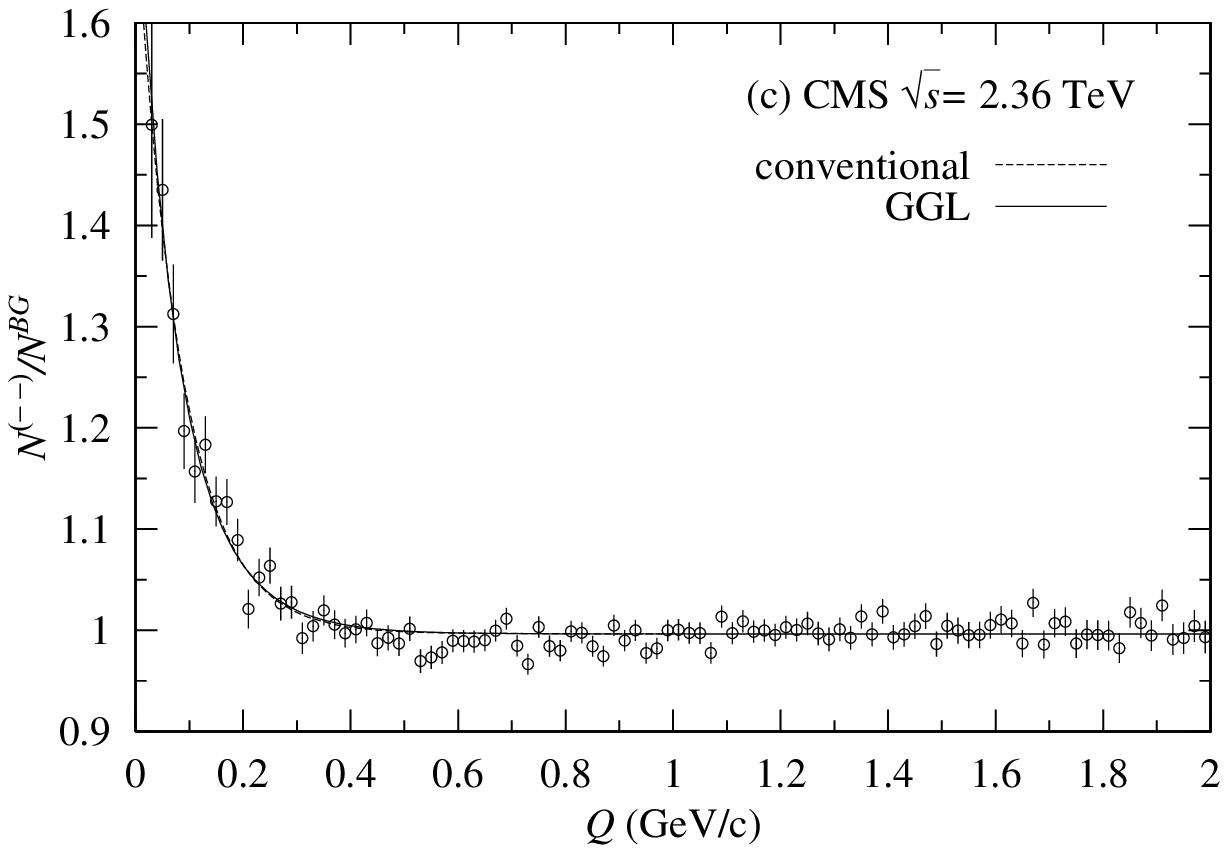}}\\
\resizebox{0.40\textwidth}{!}{\includegraphics{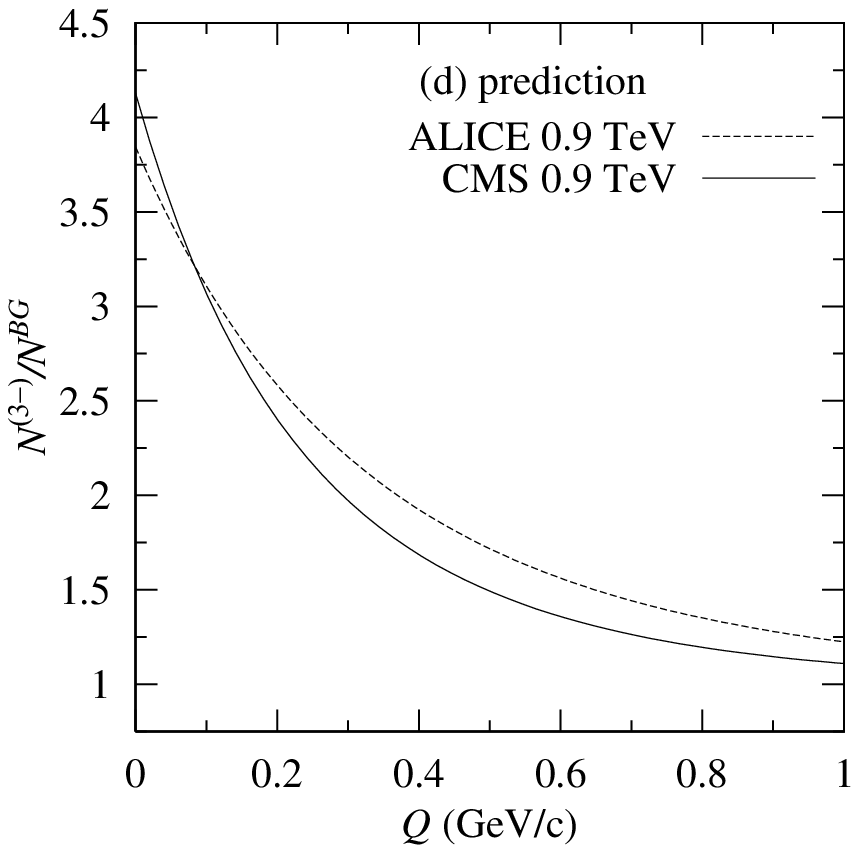}}
\resizebox{0.40\textwidth}{!}{\includegraphics{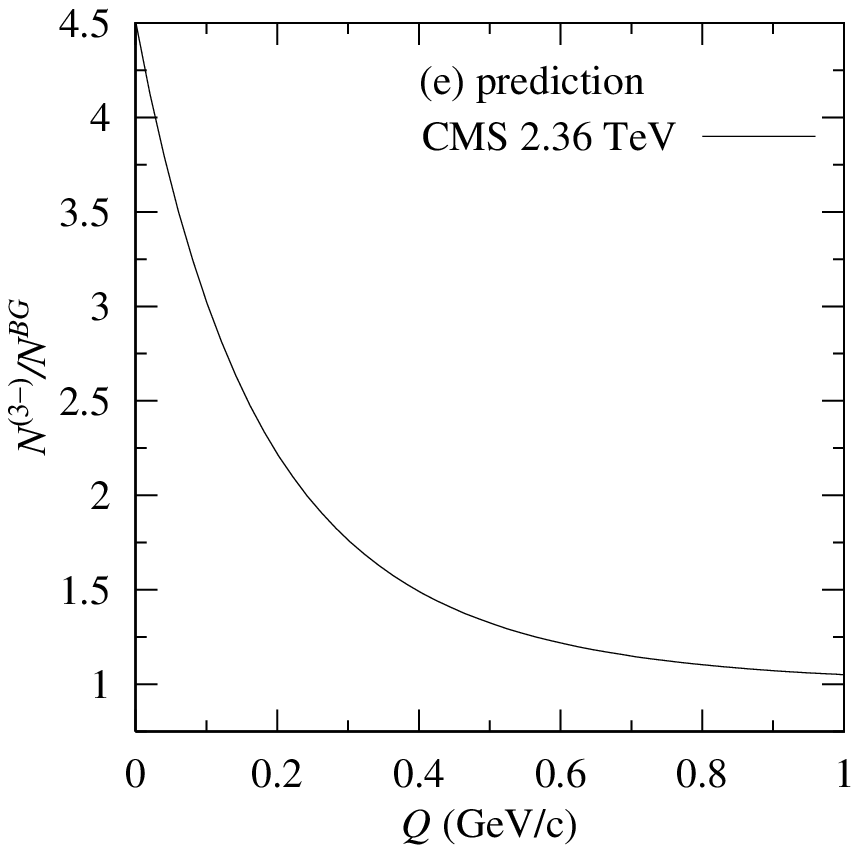}}
\caption{Analysis of data on BEC at 0.9 TeV by ALICE Collaboration with conditions $M\le$ 6, and 0.1 $\le k_T \le$ 0.55 GeV and at 0.9 and 2.36 TeV by CMS Collaboration. Panels (d) and (e) are also our predictions of the 3rd order BEC for at 0.9 and 2.36 Te.}
\label{fi_04}
\efw

\btw
\label{ta_04}
\caption{Analysis of data on BEC by ALICE Collaboration and CMS Collaboration. Because estimated value of $1/k^{(\rm BEC)}$ is a unit, it is not cited.}
\vspace{2mm}
\begin{tabular}{cccc|cccc}
\hline
\multicolumn{4}{c|}{Eq.~(\ref{eq_13})} & \multicolumn{4}{c}{Eq.~(\ref{eq_14})}\\
\multicolumn{8}{c}{(upper: Gaussian formula, and lower: exponential formula)}\\
$\lambda$ & $c$ & $R$ (fm) & $\chi^2/$NDF & 
$\gamma$  & $c$ & $R$ (fm) & $\chi^2/$NDF\\
\hline
\multicolumn{8}{c}{$\sqrt{s} =$ 0.9 TeV, ALICE (multiplicity $M\le$ 6, 0.1 $\le k_T \le$ 0.55 GeV)}\\
\hline
0.35$\pm$0.02 & 0.99$\pm$0.00 & 0.83$\pm$0.04 & 121/72 & 
4.0$\pm$ 0.3  & 0.99$\pm$0.00 & 0.81$\pm$0.03 & 119/72\\
0.64$\pm$0.04 & 0.98$\pm$0.00 & 1.33$\pm$0.09 & 98/72 &
1.30$\pm$0.23 & 0.98$\pm$0.00 & 1.18$\pm$0.07 & 98/72\\
\hline
\multicolumn{8}{c}{$\sqrt{s} =$ 0.9 TeV, CMS (Excluding 0.6 $<Q<$ 0.9 GeV/$c$)}\\
\hline
0.32$\pm$0.01 & 0.99$\pm$0.00 & 0.96$\pm$0.02 & 407/165 & 
4.5$\pm$ 0.2  & 0.99$\pm$0.00 & 0.95$\pm$0.02 & 394/165\\
0.66$\pm$0.02 & 0.99$\pm$0.00 & 1.75$\pm$0.04 & 229/165 & 
1.08$\pm$0.13 & 0.99$\pm$0.00 & 1.59$\pm$0.03 & 225/165\\
\hline
\multicolumn{8}{c}{$\sqrt{s} =$ 2.36 TeV, CMS (Excluding 0.6 $<Q<$ 0.9 GeV/$c$)}\\
\hline
0.33$\pm$0.03 & 1.00$\pm$0.00 & 1.20$\pm$0.07 & 80/81 & 
4.3$\pm$ 0.6  & 1.00$\pm$0.00 & 1.18$\pm$0.07 & 80/81\\
0.72$\pm$0.08 & 1.00$\pm$0.00 & 2.32$\pm$0.17 & 75/81 & 
0.84$\pm$0.39 & 1.00$\pm$0.00 & 2.03$\pm$0.08 & 76/81\\
\hline
\end{tabular}
\etw

\bfn
\resizebox{0.48\textwidth}{!}{\includegraphics{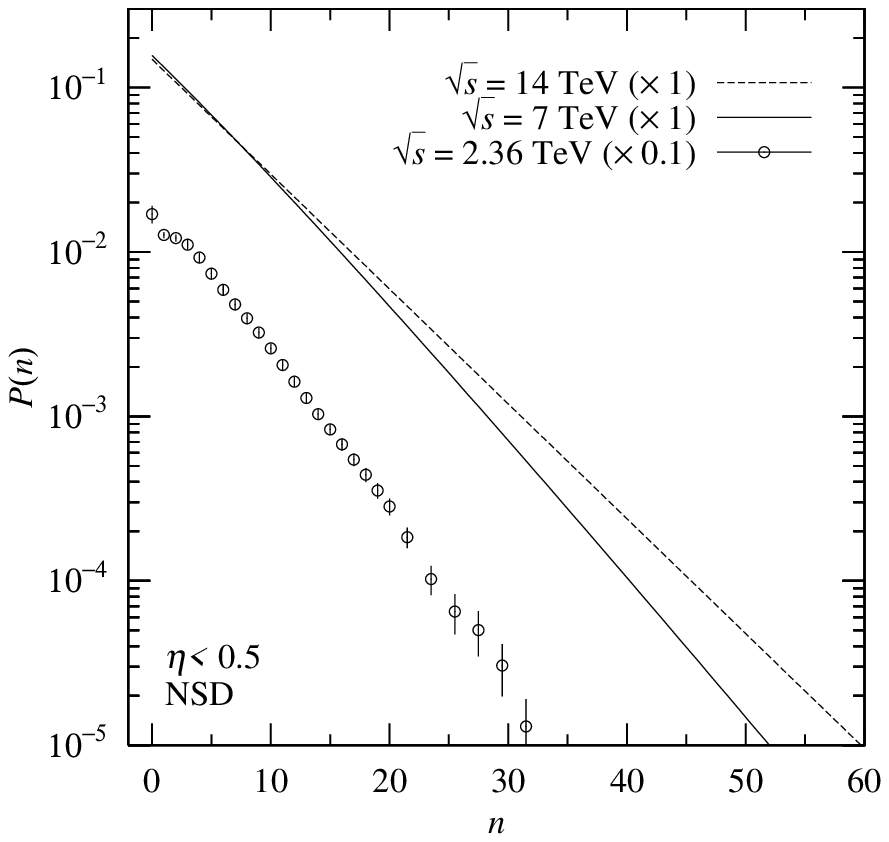}}
\caption{Expected multiplicity distributions with $\eta_c =$ 0.5 at $\sqrt{s} =$ 7 and 14 TeV. Computations are based on the GGL formula (Eq.~(\ref{eq_03})) with values in Table~\ref{ta_02} and $\av{n} = 2.5+0.76\ln(\sqrt{s}/0.2)$.}
\label{fi_05}
\efn

\section{Concluding remarks and discussion}
\label{se_05}
Through our present analyses, we summarize our concluding remarks (C1$\sim$C4), and add discussion (D1) for the information entropy.

\begin{description}
\item[C1] We have confirmed that the multiplicity distributions with $\eta_c =$ 0.5 are described by the single NBD~\cite{Aamodt:2010ft}. Moreover, we also confirm that the GGL formula does work well for the explanation of the same data in present analyses. As the pseudo-rapidity cutoffs increase, the $\eta_c =$ 1.0, and 1.3 at $\sqrt{s} =$ 0.2, 0.9 and 2.36 TeV show slightly weak violations in KNO scaling distributions, because of large values of ($\chi^2/$NDF)'s.

\item[C2] Estimated $\chi^2$ in the GGL formula are slightly better than those of the NBD. In Table~\ref{ta_02}, we observed that distributions with $\eta_c =$ 0.5 at 7 TeV does not have the coherent component. In other words, the multiplicity distributions with $\eta_c =$ 0.5 at 7 TeV are described by the NBD with $k=$ 1.

\item[C3] Using values in Table~\ref{ta_02}, we can predict multiplicity distributions with $\eta_c =$ 0.5 at 7 and 14 TeV in Fig.~\ref{fi_05}. Those are able to be examined in a near future. If there were discrepancies among data and predictions, we should consider the other effect, for example, due to the mini-jets~\cite{Giovannini:1998zb}.

\item[C4] Through present analyses of the BEC, results by the exponential formula seem to be better than those by the Gaussian formula in Table~\ref{ta_04}. See~\cite{Shimoda:1992gb} for the source functions. Moreover, values of $\gamma$'s obtained in Tables~\ref{ta_01} and \ref{ta_04} seem to be similar each other. To obtain more significant knowledge on the parameter $\gamma$, analyses of the multiplicity distributions and the BEC in the same hadronic ensembles are necessary\cite{Biyajima:1990ku,Weiner:1997kg}. Our predictions for the 3rd order BEC at $\sqrt s =$ 0.9 and 2.36 GeV would be compared by measurements as UA1 Minimum Bias Collaboration did~\cite{Neumeister:1991bq}. By the comparisons, we could obtain more useful information on the parameter $\gamma$ and the role of the GGL formula.\\

\item[D1] We have to consider reasons of the large value of $\chi^2$ concerning data with $\eta_c = 0.5$ at 0.54 TeV in Table~\ref{eq_03}. For this aim, we use the information entropy defined as~\cite{Simak:1988qp},
\be
S = -\sum_n P(n)\ln P(n).
\label{eq_16}
\ee
Results are depicted in Table~\ref{ta_05}. From them, we calculate a new plot of $(\eta_c/Y_{max},\ S/Y_{max})$, where $Y_{max} = \ln \sqrt{s}/m_p$. We see a kind of scaling law on the information entropy for multiplicity distribution~\cite{Ansorge:1988kn}.

\btn 
\label{ta_05}
\caption{The information entropy of data by ALICE Collaboration and UA5 Collaboration. $\Delta S = -\sum \delta P(n)\ln P(n) -\sum \delta P(n)$. Theoretical values are $S^{\rm (NBD)} \cong S^{\rm (GGL)} \cong S \pm$ 0.03.}
\vspace{2mm}
\begin{tabular}{l|ccccc}
\hline
$\sqrt{s}$ (TeV) & $\eta_c$ & $S$ & $\Delta S$\\
\hline
\lw{0.2}       & 0.5 & 2.05 & 0.10\\
               & 1.5 & 3.03 & 0.25\\
\hline
\lw{0.54}      & 0.5 & 2.23 & 0.07\\
               & 1.5 & 3.22 & 0.13\\
\hline
\lw{0.9 (UA5)} & 0.5 & 2.37 & 0.13\\
               & 1.5 & 3.37 & 0.29\\
\hline
               & 0.5 & 2.39 & 0.11\\
0.9 (ALICE)    & 1.0 & 3.03 & 0.18\\
               & 1.3 & 3.27 & 0.22\\
\hline
               & 0.5 & 2.56 & 0.15\\
2.36           & 1.0 & 3.22 & 0.25\\
               & 1.3 & 3.41 & 0.29\\
\hline
7              & 1.0 & 3.45 & 0.08\\
\hline
\end{tabular}
\etn

\bfn
\resizebox{0.48\textwidth}{!}{\includegraphics{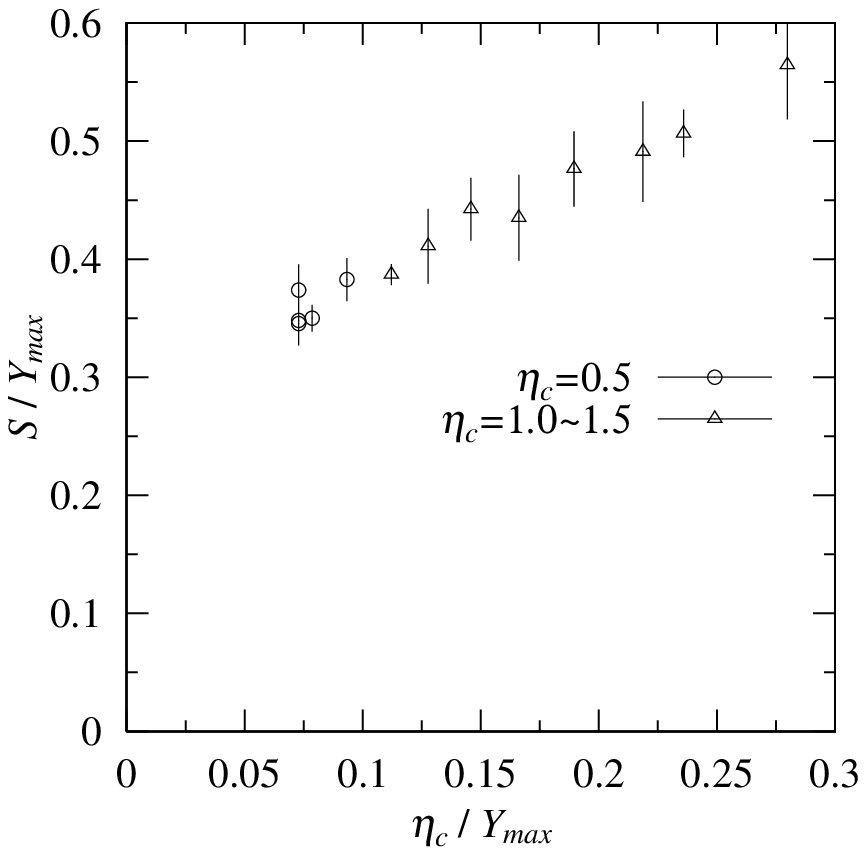}}
\caption{Distributions of data sets $(\eta_c/Y_{max},\ S/Y_{max})$, where $Y_{max} = \ln \sqrt{s}/m_p$. See Ref.~\cite{Ansorge:1988kn}}
\label{fi_06}
\efn

Next, we consider the information entropy in the KNO scaling, which is calculated by the following formula, 
\be
S^{\rm (KNO)} - \ln \av{n} = - \int \psi (z) \ln  \psi (z) dz,
\label{eq_17}
\ee
where $z = n/\av{n}$. Our results at 0.54 and 0.9 TeV are given in Table~\ref{ta_06}, where theoretical $S$'s are given as $S^{({\rm Eq.}(\ref{eq_02}))}$ and $S^{({\rm Eq.}(\ref{eq_04}))}$. We see that the differences $\delta S = S^{\rm (data)} - S^{\rm (theory)}$ between data with $\eta_c =$ 0.5 at 0.54 and 0.9 GeV and theoretical values are larger than those of other cases. The large values of the ratio $\chi^2$/NDF at 0.54 and 0.9 TeV in Table~\ref{ta_02} on KNO scaling are also observed as the large values of $\delta S$'s with $\eta_c =$ 0.5.

\btn 
\caption{The information entropy of KNO scaling distributions at $\sqrt{s} =$ 0.54 and 0.9 TeV. $S^{({\rm Eq.}(\ref{eq_02}))}$ and $S^{({\rm Eq.}(\ref{eq_04}))}$ are calculated by the use of Eq.~(\ref{eq_17}) and estimated values in Table~\ref{ta_03}.}
\vspace{2mm}
\begin{tabular}{l|ccccc}
\hline
$\sqrt{s}$ (TeV) & $\eta_c$ & $S^{\rm (KNO)}$ & $S^{({\rm Eq.}(\ref{eq_02}))}$ & $S^{({\rm Eq.}(\ref{eq_04}))}$\\
\hline
\lw{0.54}   & 0.5 & 2.23 & 2.09 & 2.09 \\
            & 1.5 & 3.23 & 3.17 & 3.18 \\
\hline
            & 0.5 & 2.39 & 2.27 & 2.28 \\
0.9 (ALICE) & 1.0 & 3.04 & 2.97 & 2.98 \\
            & 1.3 & 3.29 & 3.23 & 3.25 \\
\hline
\end{tabular}
\label{ta_06}
\etn

\end{description}

\end{document}